\documentclass[aps,preprint,11pt,nofootinbib,showpacs,showkeys]{revtex4}%
\usepackage{amsfonts}
\usepackage{amsmath}
\usepackage{amssymb}
\usepackage{graphicx}
\usepackage{color}

\begin{document}
\title[How can one probe Podolsky Electrodynamics?]{How can one probe Podolsky Electrodynamics?} \thanks{Corresponding author: C. A. M. de Melo; email: cassius.anderson@gmail.com ; tel +55 35 99586714; address: R. Corumb\'{a}, 72 - Jardim dos Estados, CEP 37701-100, Po\c{c}os de Caldas, MG, Brazil.}
\author{R. R. Cuzinatto$^{1,2}$}\thanks{rcuzin@phys.ualberta.ca; cuzinatto@unifal-mg.edu.br}
\author{C. A. M. de Melo$^{2,3}$}\thanks{cassius@unifal-mg.edu.br}
\author{L. G. Medeiros$^{4}$}\thanks{leogmedeiros@ect.ufrn.br}
\author{P. J. Pompeia$^{1,5}$}\thanks{pompeia@ift.unesp.br. On leave from:
Instituto de F\'{\i}sica Te\'{o}rica, UNESP, S\~{a}o Paulo, Brazil.}
\affiliation{$^{1}$Theoretical Physics Institute, University of Alberta, Edmonton, Alberta,
Canada, T6G 2J1.}
\affiliation{$^{2}$Instituto de F\'{\i}sica Te\'{o}rica, Universidade Estadual Paulista.}
\affiliation{R. Dr. Bento Teobaldo Ferraz, 271 Bloco II - Barra Funda,
P.O.Box 70532-2, CEP 01156-970, S\~{a}o Paulo, SP, Brazil.}
\affiliation{$^{3}$Universidade Federal de Alfenas, Campus Po\c{c}os de Caldas, Departamento de Ci\^{e}ncia e Tecnologia,}
\affiliation{R. Corumb\'{a} 72 - Jardim dos Estados, CEP 37701-100,
Po\c{c}os de Caldas, MG, Brazil }
\affiliation{$^{4}$Escola de Ci\^{e}ncia e Tecnologia, Universidade Federal do Rio Grande do Norte.}
\affiliation{Av. Hermes da Fonseca 1111 - Tirol,  CEP 59020-315, Natal, RN, Brazil}
\affiliation{$^{5}$Comando-Geral de Tecnologia Aeroespacial, Instituto de Fomento e
Coordena\c{c}\~{a}o Industrial.}
\affiliation{Pra\c{c}a Mal. Eduardo Gomes 50, CEP 12228-901, S\~{a}o Jos\'{e} dos Campos,
SP, Brazil.}

\begin{abstract}
We investigate the possibility of detecting the Podolsky generalized
electrodynamics constant $a$.
First we analyze an ion interferometry apparatus proposed by B. Neyenhuis, et al (Phys. Rev. Lett. 99,
(2007) 200401) who
looked for deviations from Coulomb's inverse-square law in the context of Proca model.
Our results show that this
experiment has not enough precision for measurements of $a$. In order to
set up bounds for $a$ we investigate the influence of Podolsky's
electrostatic potential on the ground state of the Hydrogen atom. The value of the
ground state energy of the Hydrogen atom requires Podolsky's constant to be
smaller than $5.6$ fm, or in energy scales larger than $35.51$ MeV.

\end{abstract}


\maketitle

%\volumeyear{2008} \volumenumber{number} \issuenumber{number}
%\eid{identifier}
%\date[Date text]{date}
%\received[Received text]{date} \revised[Revised text]{date}
%\accepted[Accepted text]{date} \published[Published text]{date}
%\startpage{1}
%\endpage{}

\section{Introduction\label{sec-Intro}}

The inference of the mass of the particles is a key problem in
Physics. The Higgs mechanism is the most simple and popular way to
generate massive particles from an originally gauge invariant
massless theory. From the theoretical point of view the existence of
a massive photon, usually considered in the context of Proca model,
has many implications. One of the most important is the fact that
interactions between particles are commonly described in terms of
gauge theories and, as it is well known, the gauge field is supposed
to be massless \cite{Utiyama}.
Since the electromagnetic interactions are described in terms of the $U\left(
1\right) $ symmetry group, all Quantum Electrodynamics, which is
constructed on a gauge framework, should be reviewed if a mass for the photon was verified.
The same occurs for instance in Atomic Physics, where the energy spectrum is supposed to
be different if a non-Coulomb potential is considered.

Although it is widely accepted by physicists (especially by the theoreticians)
that the photon is a massless particle, this is not an affirmation that can be
easily done from the experimental point of view since all experiments are
subject to uncertainties -- the experimentalists basically establish upper
limits for the photon mass.

Many experiments have been proposed to measure the mass of the photon
{\cite{MassTest}} and among them, several try to accomplish this by using the fact
that the electric field produced by a point charge is not the one
predicted by Coulomb law if the photon is supposed to be massive. They try to
verify the existence of a photon mass by looking for small deviations from the
Coulomb law \cite{Ref11PRL} -- usually a potential $1/r^{1+\delta}$ is tested,
and $\delta$ is evaluated. However, as mentioned in \cite{PRL}, the problem
with this type of potential is that it does not come from any underlying
theory and usually many assumptions regarding the measurement of $\delta$ are
done, so that its evaluation is strongly dependent on these hypothesis. In
order to avoid these problems the authors of {\cite{PRL}} proposed an
experiment where an ion interferometry is used to measure the photon mass. The
idea of the experiment consists, roughly speaking, in using interferometry of
an ion beam that passes through a tube where different voltages are applied --
if the mass of the photon is non-null then a difference in the interferometer
phase is expected. According to the authors of \cite{PRL}, the experiment will
be very accurate, predicting a sensitivity to the (Proca) mass of
$9\times10^{-50}~$g, \textquotedblleft2 orders of magnitude smaller than the
limit in \cite{Ref14PRL}\textquotedblright. In their case the underlying
theory is the Proca model.

However, if instead of using the Proca model, the Podolsky Generalized
Electrodynamics \cite{Podolsky} is taken into account, it is still possible to
find a mass for the (massive mode of the) photon and preserve gauge symmetry. In a recent paper
\cite{NoisNaAnnals}, a gauge theory for systems depending on the second order
derivative of the gauge field was developed and it was verified that the gauge
Lagrangian should depend on the usual field strength, $F_{\mu\nu}^{a}$, and on
its covariant derivative, $G_{\rho\mu\nu}^{a}=D_{\rho b}^{a}F_{\mu\nu}^{b}$.
In particular, for the $U\left(  1\right)  $ group it was verified that the
Podolsky Lagrangian\footnote{To preserve the correct units of the Lagrangian,
the constant $a$, henceforth referred as the Podolsky constant, has unit
$\frac{1}{l}$, where $l$ stands for length; the metric signature $\left(
+---\right)  $ is considered.},
\[
L=-\frac{1}{4}F_{\mu\nu}F^{\mu\nu}+\frac{a^{2}}{2}\partial_{\rho}F^{\rho\mu
}\partial_{\sigma}F_{~\mu}^{\sigma},\qquad F_{\mu\nu}=\partial_{\mu}A_{\nu
}-\partial_{\nu}A_{\mu},
\]
fulfills all the requirements of a second order gauge theory with an important
feature: all Lagrangians of the type $G^{2}$ for the $U\left(  1\right)  $
group differs from Podolsky Lagrangian only by a total divergence. The
(fourth-order) field equations obtained from this Lagrangian are
\[
\left(  1+a^{2}\square\right)  \partial_{\mu}F^{\nu\mu}=0,
\]
and under a generalized Lorenz condition \cite{Pimentel}, $\left(
1+a^{2}\square\right)  \partial_{\mu}A^{\mu}\left(  x\right)  =0$,
massive and massless modes for $A_{\mu}$ are identified:
\[
p^{2}\left(  1-a^{2}p^{2}\right)  A_{\mu}\left(  p\right)  =0\Rightarrow
\left\{
\begin{array}
[c]{l}%
p^{2}=0\\
p^{2}-\frac{1}{a^{2}}=0
\end{array}
\right.  .
\]

The massless mode should be understood as the
usual photon, while the massive mode was tentatively interpreted by Podolsky as
being a neutrino. This interpretation is of course outdated.  Since its original formulation,
several aspects of this theory have been analyzed, including its canonical
structure {\cite{Pimentel,Pod_Const}}, quantization {\cite{Pod_Quant}},
and others {\cite{Pod_Others}}. Several problems of this theory have been pointed out,
such as unitarity violation and the presence of ghost
states with negative norm, typical of theories with higher derivatives {\cite{PaisUhlenbeck}},
but on the other hand good properties were also obtained (see references in {\cite{Pod_Quant}}),
what motivates the study of systems of this kind nowadays, specially in the context
of an effective field theory (EFT). It is as an EFT that Podolsky theory should be understood
and in this sense the parameter $a$ sets the length scale where
the theory is valid. We also emphasize that only
classical aspects of the Podolsky theory will be considered,
so that some problems typical of the quantization procedure should
not be a concern here.

Since Podolsky electrodynamics predicts the existence of a massive mode
for the photon, if the experiment proposed in {\cite{PRL}} finds a
deviation in the interferometer phase, then this could be either an
indicative of the existence of the photon mass in the context of the
Proca model or of the existence of a non-null
value for Podolsky constant, giving support to the Podolsky theory.
One of the purposes of the present work is to analyze how the
Podolsky constant can be determined or constrained by the ion
interferometry experiment proposed in {\cite{PRL}}. This is
discussed in Section \ref{sec-Ion}, where the analytical solution
for the problem will be analyzed and numerical estimations for
Podolsky constant will be made.

On the other hand, if Podolsky
theory is to be verified, then many implications in other known
results are expected. As an example, the energy spectrum of the
Hydrogen atom as described by Quantum Mechanics is to be altered,
since the Coulomb potential should be substituted by the potential
predicted by Podolsky Electrodynamics. This is the second point to
be studied here. A perturbative solution for the Quantum Mechanics
wave function of the electron will be obtained -- see Section
\ref{sec-H} -- and another constraint on $a$ will be made. Section
\ref{sec-Conclusion} presents our conclusions.

\section{Ion Interferometry Experiment \label{sec-Ion}}

In the experiment proposed in {\cite{PRL}} a time-varying voltage is applied
to a conducting cylinder that is nested inside a grounded second cylinder. A
beam of ions pass through the inner conductor through three gratings, forming
a Mach-Zehnder interferometer -- for more details see the original paper. If
there is an electric field inside the cylinder, i.e. if the ions go through
different potentials, then an interferometer phase shift is expected. Notice
that this is not what is predicted by Maxwell equations for a conducting
shell, according to which the potential inside the apparatus should be constant.

After passing through the first grating the ion beam is split in two arms: one
travels horizontally (parallel to the cylinder axis), while the second goes
diagonally. When the two arms reach the second grating, the one that was
advancing horizontally begins to travel diagonally while the second starts to
go horizontally, until they reach the third grating, where they become one
single beam travelling horizontally. Since the distance between the gratings is
the same, the diagonal segments of each arm travel through the same potentials
and they induce the same phase shift. However, the segments of the arms that
go horizontally pass through different potentials; if there is a phase
shift in the interferometer it is caused by the difference of potentials
between the horizontal segments (see Fig. 1). We consider that
the distances of the horizontal segments from the center of the cylinder are
$r_{0}$ and $r_{0}+s$. This way, what the interferometer actually does is to
measure a phase shift induced by the potential difference between these
horizontal segments of arms the split beam.

The first information required is the equation for the potential inside the
cylinder as predicted by the theory. In {\cite{PRL}} the authors considered
the Proca model. Here we will analyze Podolsky Electrodynamics \cite{Podolsky}%
, where the equation for the electrostatic potential is given by
\[
\left(  1-a^{2}\nabla^{2}\right)  \nabla^{2}\phi=0.
\]
To solve this equation, let us define
\[
U\equiv\nabla^{2}\phi.
\]
First we solve the homogeneous equation for $U$
\begin{equation}
\left(  1-a^{2}\nabla^{2}\right)  U=0, \label{EqU}%
\end{equation}
and then consider the non-homogeneous equation for $\phi$,
\begin{equation}
\nabla^{2}\phi=U_{h}, \label{EqPhi}%
\end{equation}
where $U_{h}$ is a solution of (\ref{EqU}). In view of the symmetry of the
problem, cylindrical coordinates are considered and no angular dependence is
expected. Also, since the inner cylinder has an elongated geometry, the
infinite tube approximation can be done and no longitudinal dependence exists.
The solution for (\ref{EqU}) is found under these assumptions, and Eq.
(\ref{EqPhi}) becomes
\begin{equation}
\frac{1}{r}\frac{d}{dr}\left(  r\frac{d\phi}{dr}\right)  =AI_{0}\left(
\frac{r}{a}\right)  +BK_{0}\left(  \frac{r}{a}\right)  , \label{DiffEqPhi}%
\end{equation}
where $I_{0}$ and $K_{0}$ are the modified Bessel functions of the first and
second kind.

The integration of Eq. (\ref{DiffEqPhi}) gives us
\begin{equation}
\phi\left(  \frac{r}{a}\right)  =a^{2}AI_{0}\left(  \frac{r}{a}\right)
+a^{2}BK_{0}\left(  \frac{r}{a}\right)  +D\ln\frac{r}{a}+C. \label{Solution}%
\end{equation}
This solution carries a desirable feature: the homogeneous part is the usual
Maxwell term and the particular solution is the Podolsky contribution. In
fact, this split always occurs in the electrostatic case of Podoslky theory
when vacuum is assumed.

Four integration constants appear in the solution (\ref{Solution}), as
expected from a fourth-order equation, and boundary conditions are
used to fix them. First we consider that the potential in the limit
$r\rightarrow0$ should be
finite. Using the asymptotic form for $I_{0}$ and $K_{0}$ \cite{Butkov,Arfken}%
{,} we conclude that $D=a^{2}B$. Another boundary condition that is
used is the value of the potential at $r=R$, where $R$ is the radius
of the inner tube. If $V_{0}$ is the voltage applied to the inner
tube relative to the outer tube, whose unknown (ground) potential is
$V_{g}$, then
\[
V_{0}+V_{g}=a^{2}A\left[  I_{0}\left(  \frac{R}{a}\right)  +g\left(  a\right)
\left[  K_{0}\left(  \frac{R}{a}\right)  +\ln\frac{R}{a}\right]  +f\left(
a\right)  \right]  ,
\]
where $B$ and $C$ were redefined as $B=g\left(  a\right)  A$ and
$C=f\left( a\right)  a^{2}A$, and $A$ is supposed to be non-null.
This expression is used to determine $A$ in terms of the other
constants. Yet another expected boundary condition is that the
electric field $\mathbf{E}$ at $r=0$ is null (otherwise it would be
discontinuous without a physical reason). Actually with the
redefinitions of $B$ and $C$ above, it is verified that this
condition is already satisfied, so that no other constant is fixed
with this condition. However, if we claim that the
divergent of the electric field is
finite at
$r=0$,\footnote{What makes the electric
field flux finite at the origin.} then we must set $g\left(
a\right) =0$. At last, in order to fix $f\left( a\right) $ we assume
that the potential at $r=0$ can be measured -- this is an additional
step in the experimental procedure proposed in \cite{PRL} where no
measurement of the potential at $r=0$ is suggested; in our case this
is essential for determining the last integration constant. We
suppose that the measured $\phi\left( 0\right)  $ is expressed as
$\phi\left(  0\right) =\left( V_{0}+V_{g}\right) \epsilon$, with
$0\leq\epsilon\leq1$. This fixes $f\left(  a\right) $ as
\[
f\left(  a\right)  =\frac{\epsilon I_{0}\left(  \frac{R}{a}\right)  }{\left(
1-\epsilon\right)  }.
\]

Finally the potential is written as
\[
\phi\left(  \frac{r}{a}\right)  =\left(  V_{0}+V_{g}\right)  \left[
\frac{I_{0}\left(  \frac{r}{a}\right)  }{I_{0}\left(  \frac{R}{a}\right)
}\left(  1-\epsilon\right)  +\epsilon\right]  .
\]
Notice that if no Podolsky term is supposed to exist, then the potential
inside the inner tube will be the same everywhere, i.e. $V_{0}+V_{g}$, which
means that $\epsilon=1$.

Now the potential difference between the horizontal segments of the
arms of the split beam can be evaluated as
\[
\Delta\phi=\phi\left(  \frac{r_{0}+s}{a}\right)  -\phi\left(  \frac{r_{0}}%
{a}\right)  =\left(  V_{0}+V_{g}\right)  \left[  \frac{I_{0}\left(
\frac{r_{0}+s}{a}\right)  -I_{0}\left(  \frac{r_{0}}{a}\right)  }{I_{0}\left(
\frac{R}{a}\right)  }\right]  \left(  1-\epsilon\right)  .
\]

The interferometer phase is given by
\[
\Phi=\frac{e\tau}{\hbar}\Delta\phi+\Phi_{0}=\frac{e\tau}{\hbar}\left(
V_{0}+V_{g}\right)  \left[  \frac{I_{0}\left(  \frac{r_{0}+s}{a}\right)
-I_{0}\left(  \frac{r_{0}}{a}\right)  }{I_{0}\left(  \frac{R}{a}\right)
}\right]  \left(  1-\epsilon\right)  +\Phi_{0},
\]
where $e$ is the charge of the ion (in the present case $e$ is the electron
charge), $\tau$ is the time that the ion takes to travel lengths of the
horizontal segments and $\Phi_{0}$ is the phase indicated by the
interferometer when $V_{0}+V_{g}=0$. In order to eliminate the two unknown
constants $\Phi_{0}$ and $V_{g}$, two potential differences $V_{0}$ and
$V_{0}+\Delta V$ can be applied to the inner tube. The difference in the
phases due to this change will be
\[
\Delta\Phi=\frac{e\tau}{\hbar}\Delta V\left[  \frac{I_{0}\left(  \frac
{r_{0}+s}{a}\right)  -I_{0}\left(  \frac{r_{0}}{a}\right)  }{I_{0}\left(
\frac{R}{a}\right)  }\right]  \left(  1-\epsilon\right)  .
\]
This expression is inverted in order to obtain the Podolsky constant
$a$ as a function of the experimental parameters. This will be done
under some assumptions. First we expect that the value of Podolsky
constant is small, so that only small differences from Maxwell
equations can be detected. If this is the case, then the asymptotic
limit for $I_{0}$ can be used
\cite{Butkov,Arfken}, $I_{0}\left(  x\right)  \sim\frac{1}{\sqrt{2\pi x}}%
e^{x}$. This allows us to estimate the Podolsky constant as
\begin{equation}
a=\frac{R-\left(  r_{0}+s\right)  }{\ln\left(  1-\epsilon\right)  -\ln\left(
\frac{\hbar}{e\tau}\frac{\Delta\Phi}{\Delta V}\sqrt{\frac{r_{0}+s}{R}}\right)
}. \label{a}%
\end{equation}
Notice that $\lim_{\epsilon\rightarrow1}a=0$, which means that Electrodynamics
reduces to the Maxwell one.

We shall obtain numerical estimations for $a$ considering ion beams
composed by $^{1}H^{+}$ and $^{133}Cs^{+}$. According to \cite{PRL},
these ions can travel at a speed $v$ of $311\,m/s$ and $27\,m/s$
respectively; the length of the horizontal segments are fixed in
$1\,m$ so that $\tau=L/v$ is determined for both cases. The
potential difference $\Delta V$ can be fixed as $400\,kV$ and the
values of $R,\,r_{0}$ and $s$ are set to $R=27\,cm$, $r_{0}\left(
^{1}H^{+}\right)  =24.4\,cm$, $r_{0}\left(  ^{133}Cs^{+}\right)
=24.9\,cm$, and $s\left(  ^{1}H^{+}\right)  =6.4\,mm$, $s\left(
^{133}Cs^{+}\right) =0.56\,mm$. Fig. 2 shows the numerical
estimations for $a$ for different values of $\epsilon$ and
$\Delta\Phi$ for $^{1}H^{+}$. The range of values for $\epsilon$ was
established considering the fact that a precision of $10^{-8}$ could
be achieved with the available commercial multimeters (in the best
case). Concerning $\Delta\Phi$, it was considered that phase shifts
as small as $10^{-4}\,\text{rad}$ can be detected \cite{PRL}.

According to these numerical evaluations, the experiment would be able to
detect values of the Podolsky constant $a_{Cs^{+}}\geq0.033~cm$ in the
case of $^{133}Cs^{+}$ ion beam and $a_{H^{+}}\geq0.069~cm$ if the
$^{1}H^{+}$ ion beam is used. These values seem consistent with the asymptotic
limit taken for $I_{0}$; indeed, they are small when compared to the values of
$R$ and $r_{0}$ and therefore the ratios that appear in $I_{0}$ -- namely,
$\frac{R}{a_{H^{+}}}=391.3,~\frac{R}{a_{Cs^{+}}}=810.81,~\frac{r_{0}}%
{a_{H^{+}}}=353.62,~\frac{r_{0}}{a_{Cs^{+}}}=732.73$ -- are all of order of
$10^{2}-10^{3}$.

The mass of the photon is evaluated using these values for $a$ and
the expression
\begin{equation}
m_{\gamma}=\frac{\hbar}{ac}. \label{mass}%
\end{equation}
As the mass scales with the inverse of the Podolsky constant, the smallest
value of $a$ that can be measured will give the greatest measurable value for
the photon mass. Each ion beam will predict a different upper limit:
$m_{\gamma}^{^{133}Cs^{+}}=1.06\times10^{-39}~kg=5.98\times10^{-8}~eV$ and
$m_{\gamma}^{^{1}H^{+}}=5.10\times10^{-40}~kg=2.85\times10^{-8}~eV$.

Although Proca and Podolsky approaches predict a massive mode for the photon,
there is some important difference between them. First, Podolsky
Electrodynamics is a gauge theory, while Proca model explicitly break such
symmetry, what could have implications for the charge conservation. Second, in
the Proca context it is expected that the photon mass, if it exist, should be
very small. Conversely, the Podolsky's massive model would be very large once
it is the inverse of the scale of length where the generalized theory is
effective, cf. Eq. (\ref{mass}). That is, Proca (Podolsky) model predicts
deviations from Maxwell electrodynamics in very low (high) energy regimes.

It is important to emphasize that the photon mass in independent of the nature
of the ion composing the beam in the experiment. The different values for
$m_{\gamma}$ for $^{133}Cs^{+}$ and $^{1}H^{+}$ express only the different
values of $a$ accessed by the experiment.

One could argue that the values of $a$ that can be measured by the ion
interferometer are very high in absolute terms. In fact, one would say that if
$a$ were of order of $10^{-2}$ as indicated here, the deviations from the
Maxwellian electromagnetism would have been detected long ago. In face of
this, the conclusion would be that the experiment proposed in \cite{PRL} is
not appropriate for measuring the Podolsky constant and therefore the photon
mass in this theory. This is indeed a strong argument, but we would like to
give a quantitative justificative for ruling out the ion beam apparatus as an
appropriate set to find the Podolsky mass.

In the next section we will make the hypothesis that Podolsky electrodynamics
hold at the atomic scale\footnote{This is not mandatory once Podolsky's theory
for the electromagnetism is an effective theory.} and see the implications for
the elementary physics of the Hydrogen atom; in particular, we will analyze
the energy of the fundamental state.

\section{Hydrogen Atom \label{sec-H}}

Now we turn to the problem of considering the Hydrogen atom, as treated by
Quantum Mechanics, where the electromagnetic potential is the one described by
Podolsky Electrodynamics. The goal of this section is to analyze the effects
of a non-null Podolsky constant and verify what are the implications of the
values found for $a$. We consider $\hbar=1$ to simplify the notation, but the
units are restored when numerical evaluations are done.

The electrostatic potential is given by \cite{Podolsky,NoisNaAnnals}
\[
\phi\left(  r\right)  =-\frac{e}{r}\left(  1-e^{-\frac{r}{a}}\right)  ,
\]
and the Hamiltonian operator reads $\hat{H}=\frac{\hat{p}^{2}}{2m}%
+e\phi\left(  r\right)  $. The variational method will be employed so that a
perturbative solution for the wave function of the ground state, $\psi\left(
r\right)  $ may be found. The tentative solution is
\[
\psi\left(  r\right)  =Ne^{-\gamma r},
\]
where $N$ is a normalization constant set as $N^{2}=\frac{\gamma^{3}}{\pi}$;
$\gamma$ is a parameter that will be determined by the variational method,
according to which the energy, given by
\[
E=\int dV\psi^{\ast}\left(  r\right)  \hat{H}\psi\left(  r\right)
=\frac{\gamma^{2}}{2m}-e^{2}\gamma+e^{2}\frac{4\gamma^{3}}{\left(
2\gamma+\frac{1}{a}\right)  ^{2}},
\]
should be minimized:
\begin{equation}
\frac{\partial E}{\partial\gamma} = \frac{8a^{3}}{m}\gamma^{4}+\frac{12a^{2}%
}{m}\gamma^{3}+\frac{6a}{m}\gamma^{2}-6ae^{2}\gamma+\frac{\gamma}{m}-e^{2}=0.
\label{dE/dg}%
\end{equation}

Now suppose that the value of the Podolsky constant is actually small, then
Eq. (\ref{dE/dg}) can be solved considering only terms up to first order in
$a$. The solution found for $\gamma$ is $\gamma_{+}=me^{2}$ and $\gamma
_{-}=-\frac{1}{6a}$. The energies evaluated with these solutions are
\[
E\left(  \gamma_{+}\right)  =-\frac{me^{2}}{2}e^{2}\left(  1-2\left(
2mae^{2}\right)  ^{2}\right)  +O\left(  a^{3}\right)  ,\qquad E\left(
\gamma_{-}\right)  =\frac{9ame^{2}+1}{72a^{2}m}.
\]
The value of $E\left(  \gamma_{-}\right)  $ gives a positive energy and for
small values of $a$ it becomes too high, therefore this result should be
excluded. $E\left(  \gamma_{+}\right)  $ can only be calculated with a given
value of $a$, but for small $a$ it is only a perturbation on the known result
given by Quantum Mechanics, $E=-\frac{me^{2}}{2}e^{2}$. If we want to find a
value for $a$ that is compatible with the known results given in the
literature we should expect the perturbation $2\left(  2mae^{2}\right)  ^{2}$
to be smaller than the relative experimental uncertainty of the energy of the
ground state. Proceeding this way it follows
\[
a\leq\frac{r_{B}}{2}\sqrt{\frac{\sigma_{E_{0}}}{2\left\vert E_{0}\right\vert
}},
\]
where $r_{B}=\frac{1}{me^{2}}$ is the Bohr radius. Restoring the units and
using values given in the literature \cite{PDG} we should expect
\begin{equation}
a\leq5.56\,\text{fm \ \ \ \ or \ \ \ }m_{\gamma}\geq35.51\,\text{MeV}
\label{Constr_Ato_Hidr}%
\end{equation}
Clearly these values for $a$ and $m_{\gamma}$ are not compatible with the
possible values that can be found in the interferometry experiment.

\section{Conclusions\label{sec-Conclusion}}

We have discussed how the ion interferometry experiment proposed in
Ref. {\cite{PRL}} could be used to measure the value of Podolsky
constant $a$ and the massive mode of the photon in the context of
Podolsky Electrodynamics. The minimum value of $a$ that could be
detected -- $a=0.033~cm$ with the $^{133}Cs^{+}$ ion beam -- is too
large as an admissible effective scale, and would lead to a mass
$m_{\gamma}\leq1.06\times10^{-39}~kg=5.98\times 10^{-8}~eV$ for the
photon which is excluded by current experimental data {\cite{PDG}.}

We might think of improving the accuracy of the measurements of the phase shift
and/or of the potential at $r=0$ (for instance, using some better technology
in the apparatus). But the logarithmic behavior of (\ref{a}) in terms of these
quantities makes this possibility unlikely: great improvements in the
detection of $\Delta\Phi$ and $\Delta V$ would lead to small changes in $a$
[see Eq.~(\ref{a})]. Therefore, this rules out the interferometric ion beam
experiment as a suitable one for testing Podolsky Electrodynamics.

Besides gauge invariance, Podolsky Electrodynamics has another
peculiar feature that distinguishes it from the Proca field: the
smaller the characteristic constant $a$ the greater the mass
associated to the photon. Hence we are strongly constrained: the
Maxwellian electromagnetism must hold until small scales of length,
and therefore $a$ has to be small, otherwise the additional
Podolsky term in the Lagrangian for the electromagnetic field would
be significant and the resulting modifications in the ordinary
theory would be easily detected. These scales of length are set, for
instance, by the spectroscopy of Hydrogen atom. So, we tested
Podolsky's theory calculating the value of $a$ that would be
consistent with the experimental error in the energy of the
fundamental level of the Hydrogen. The result, $a\leq5.56\,fm$,
clearly shows that the ion interferometer experiment does not have
enough precision to measure a Podolsky constant that is this small.

The constraint $a\leq5.56 \, fm$ coming from Quantum Mechanics
considerations set a high energy scale for the photon mass: $m_{\gamma} > 35.51$
MeV. This way, if Podolsky model is correct, it is
expected to engender deviations from Maxwell Electrodynamics only in
high energy scales, which are accessible by particle accelerators. Therefore, the next necessary step is to investigate in more detail which kind of effects appear in QED$_4$ due to the Podolsky term.

\bigskip

\acknowledgments

RRC and PJP thank the Physics Department of University of Alberta for
providing the facilities. This work was supported by FAPERJ-Brazil grant
E-26/100.126/2008, CNPq-Brazil and NSERC-Canada.

%%%%%%%%%%%%%%%%%%%%%%%%%%%%
%%%%%%%%%% FIGURAS %%%%%%%%%%
%%%%%%%%%%%%%%%%%%%%%%%%%%%%

\begin{figure}[th]
\begin{center}
\includegraphics[width=13cm]{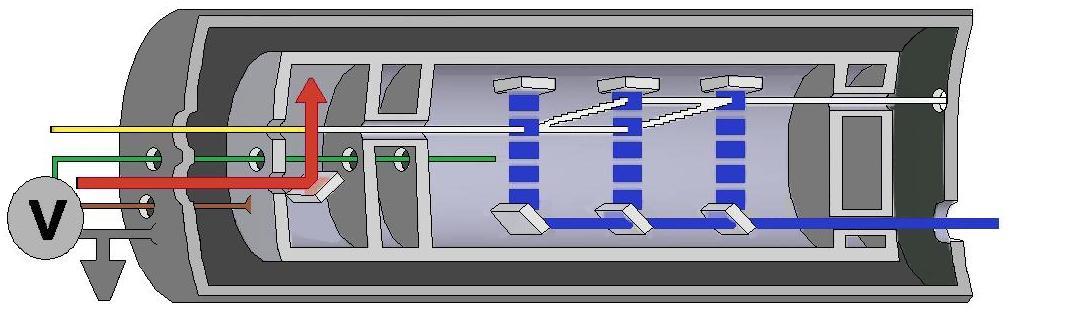}
\end{center}
\caption{Sketch of the ion interferometry experiment. A cutaway of the cylinders is shown.}%
\label{fig1}%
\end{figure}

\begin{figure}[th]
\begin{center}
\includegraphics[width=13cm]{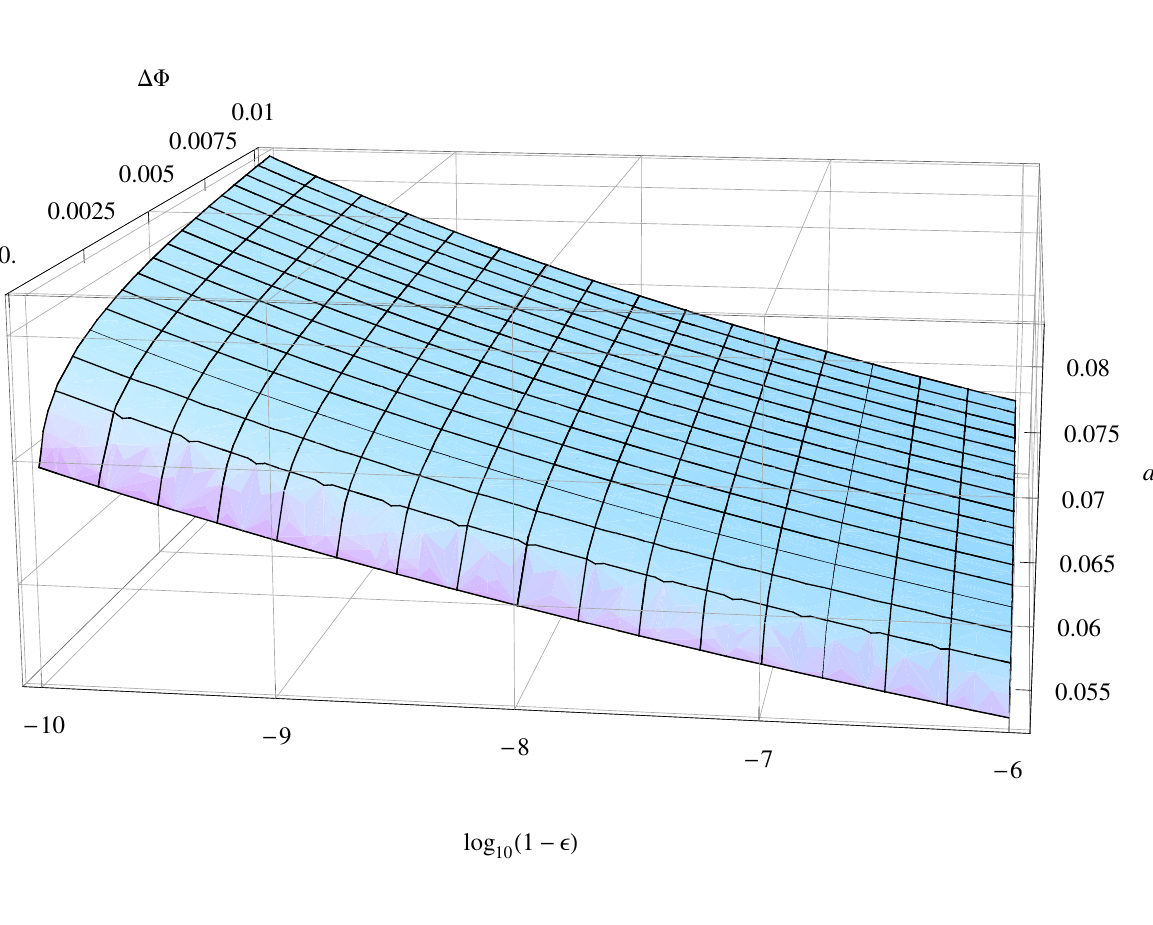}
\end{center}
\caption{Values of $a$ (cm) for different values of $\varepsilon$
(from $0.001$ to $0.999$) and $\Delta\Phi$ (rad) (from $10^{-4}$ to
$10^{-2}$) using $^{1}H^{+}$ ion beam. The graph for the $^{133}Cs^{+}$ ion beam is very similar.}%
\label{fig2}
\end{figure}
%%%%%%%%%%%%%%%%%%%%%%%%%%%%

\end{document}